%% file: Mazziotti2022Spin.tex
\RequirePackage[2020-02-02]{latexrelease}
\documentclass[twocolumn,english,prb]{revtex4-2}
\usepackage[T1]{fontenc}
\usepackage[latin9]{inputenc}
\usepackage{amsmath}
\usepackage{amssymb}
\usepackage{graphicx}
\usepackage{babel}
\usepackage{xcolor}
\usepackage{natbib}
\usepackage{listings}

\begin{document}

\title{Spin-orbit coupling controlling the superconducting dome of artificial superlattices of quantum wells}
\author{Maria Vittoria Mazziotti}
\thanks{Corresponding author. E-mail: vittoria.mazziotti@gmail.com}
\affiliation{INFN, Laboratori Nazionali di Frascati, Via Enrico Fermi 40,  Roma, Italy}
\affiliation{Rome International Center for Materials Science, Superstripes, RICMASS, Via dei Sabelli 119A, 00185 Roma, Italy}

\author{Antonio Bianconi}
\affiliation{Rome International Center for Materials Science, Superstripes RICMASS Via dei Sabelli 119A, 00185 Roma, Italy}
\affiliation{Istituto di Cristallografia - Consiglio Nazionale delle Ricerche IC-CNR, via Salaria Km 29.300, Monterotondo (Roma) I-00016, Italy}
\author{Roberto Raimondi}
\affiliation{Dipartimento di Matematica e Fisica, Università Roma Tre, via della Vasca Navale 84 00146 Roma, Italy}
\author{Gaetano Campi}
\affiliation{Istituto di Cristallografia - Consiglio Nazionale delle Ricerche IC-CNR, via Salaria Km 29. 300, Monterotondo (Roma) I-00016, Italy}
\author{Antonio Valletta}
\affiliation{Institute for Microelectronics and Microsystems IMM, Italian National Research Council CNR, via del Fosso del Cavaliere, 100, 00133 Roma, Italy}

\begin{abstract}
While it is known that a resonant amplification of $T_c$ in two-gap superconductors can be driven by using the Fano-Feshbach resonance tuning the chemical potential near a Lifshitz transition, little is known on tuning the $T_c$ resonance by cooperative interplay of the  Rashba spin-orbit coupling (RSOC) joint with phonon mediated (e-ph) pairing at selected k-space spots. Here we present first-principles quantum calculation of superconductivity in an artificial heterostructure of metallic quantum wells with 3 nm period where quantum size effects give two-gap superconductivity with RSOC controlled by the internal electric field at the interface between the nanoscale metallic layers intercalated by insulating spacer layers. The key results of this work show that fundamental quantum mechanics effects including RSCO at the nanoscale (Mazziotti et al Phys. Rev. B, 103, 024523, 2021) provide key tools in applied physics for quantitative material design of unconventional high temperature superconductors at ambient pressure. We discuss the superconducting domes where  $T_c$ is a function of either  the Lifshitz parameter ($\eta$) measuring the distance from the topological Lifshitz transition for the appearing of a new small Fermi surface due to  quantum size effects with finite spin-orbit coupling and the variable e-ph coupling $g$ in the appearing second Fermi surface linked with the energy softening of the cut off $\omega_0$.
\end{abstract}

 \date{\today}   

\maketitle

\section{Introduction}

While in the early times fundamental studies on quantum size effects in artificial superlattices (SLs) of quantum wells have been limited by technological constraints to a thickness of hundreds of nanometers [\onlinecite{dingle1974quantum}],
now the electronics industry is focusing toward devices with 3 nm technology. 
Recently, artificial superlattices of atomically thin layers down to the few nanometers scale have been
synthesized [\onlinecite{nozik2021quantization,bandurin2017high,mudd2015high,li2015quantum,mak2010atomically,baiutti2014towards,suyolcu2020design}].
The control of quantum size effects at  the nanometer level led to advances in understanding fundamental physical phenomena 
at nanoscale and to the realisation of novel functional devices. The electronic properties of these devices can be modified not only by 
a  careful selection of the materials and the thickness of quantum units within the stack, 
but also by  tuning the charge density, the spacer layers, the misfit strain and relative orientation of the layers.
Recently, there has been a growing interest in the  design of superlattices of quantum wells, where  stacks of first units   of nanometer thick superconducting layers  are 
intercalated by nanoscale insulating layers  following the European patent on  the design of 
artificial high-$T_c$  atomic-limit superconducting  heterostructures  [\onlinecite{bianconi1996high}] and 
the US patent [\onlinecite{bianconi2001process}] for  the $T_c$-resonant amplification via Fano-Feshbach resonance.
The  possibility  of   $T_c$ amplification by Fano-Feshbach 
resonance between open and close pairing channels in artificial superlattices showing two-gap superconductivity 
was confirmed by the discovery of naturally  layered two-gap superconductors like doped $MgB_2$ in 
2001,  doped iron-based superconductors in 2008,  
and oxide-oxide perovskite interfaces, which have inspired the material design of different classes of artificial nanoscale 
superlattices [\onlinecite{innocenti2010resonant,innocenti2010shape,bianconi2014shape}].
While two-gap superconductivity in the twentieth century was considered a theoretical curiosity of mathematical physics,  
 in layered systems it has become a hot topic in the first two decades of this century. 
The classical theory of two-gap superconductivity was first developed by  introducing the pair exchange between two BCS condensates 
in the weak coupling regime by Moskalenko [\onlinecite{moskalenko1959superconductivity}] and Suhl, Matthias and 
Walker (SMW), [\onlinecite{suhl1959bardeen}]. 
A second theoretical approach, the multi-band Migdal-Eliashberg theory, extended the theory to the case
where one of the two condensates is in the strong coupling regime  as was applied, for example, in
doped $MgB_2$ [\onlinecite{ummarino2004two,Bussmann2003}].
The third theoretical approach available to understand two-gap superconductivity was the mixed boson-fermion 
model [\onlinecite{friedberg1989gap}], derived from the Tomonaga theory of pions cloud in nuclear matter [\onlinecite{palumbo2016pion}] and the
the negative-U-centers model where localised bosons or bipolarons interact symbiotically with free particles in different parts 
of the unit cell,  proposed for heavy fermions
 [\onlinecite{geballe2000qualitative}] and doped
$BaBiO_3$ [\onlinecite{menushenkov2016fermi,dzero2005superconductivity}].
A fourth approach, proposed by Bianconi, Perali and Valletta 
(BPV) [\onlinecite{bianconi1998superconductivity,perali1996gap,valletta1997electronic}] 
provides the theory of two-gap superconductivity for the scenario where a first BCS pairing scattering channel (open channel) 
in a first Fermi surface resonates with a second pairing channel (closed channel) in the BCS-BEC crossover regime
in the second Fermi surface
[\onlinecite{perali2004bcs,perali2004quantitative,perali2012anomalous,shanenko2012atypical,chen2012superconducting,cariglia2016shape,guidini2014band,guidini2016bcs,doria2016multigap,salasnich2019screening,shanenko2006size,valentinis2016bcs,tajima2020investigate,vargas2020crossband,ochi2022resonant,kagan2019fermi}].

While it is known that   the BCS superconductivity theory predicts  that $T_c$ does not depend on the variation of the Fermi level,
 assumed to be very far from the band edge, in the BPV two-gap superconductors the second Fermi surface
 is at  a finite particular distance from the band edge. In  fact the  resonant $T_c$-amplification strongly depends on the relative position
 of the chemical potential  with respect  to the topological Lifshitz transition for the second Fermi surface appearance.
Therefore  $T_c$ is a function of the Lifshitz parameter $\eta$ defined as
\begin{equation}
\label{eq:7.1}
\eta=\frac{\mu-E_L}{\Delta E},
\end{equation}
where $ \mu $ is the chemical potential, $E_L$ is the second subband  bottom energy  and  $\Delta E$ is the dispersion along the  $z$ direction of the highest-energy subband. In the specific superlattice studied here $\Delta E$=30 meV.

Above the Lifshitz transition for  a new Fermi surface  appearance,  $T_c$  as a function of $\eta$
  shows a  quantum resonance between open and closed scattering channels discovered by Fano
[\onlinecite{fano1961effects}]{\color{red}.}  Such so-called Fano-Feshbach resonance  was later found
in X-ray spectroscopy [\onlinecite{cauchois1964spectroscopie,fano1968spectral,bianconi1991linearly}],
in quantum devices [\onlinecite{luk2010fano,miroshnichenko2010fano,limonov2021fano}],
and quantum optics [\onlinecite{limonov2021fano}],
shape resonance  in nuclear physics [\onlinecite{feshbach1962unified}] and in the early works in two-gap superconductivity 
[\onlinecite{bianconi1998superconductivity,perali1996gap,valletta1997electronic}], and 
Feshbach resonance in ultra cold gases [\onlinecite{zwierlein2004condensation}].
For a historical perspective see Ref.[\onlinecite{vittorini2009majorana}].
The key feature of the BPV theory is  the  calculation of the energy dependent 
pair-transfer exchange interaction, which generates the Fano-Feshbach resonance in the Bogoliubov multi-gap superconductivity theory 
[\onlinecite{bogoliubov1958new}] [\onlinecite{svistunov2015superfluid}].
The energy dependent pair-transfer exchange term is calculated from the overlap of the four wave-functions of the
 two-electron pairs at the two Fermi levels  as  obtained from the solution of the Schr\"odinger equation  
[\onlinecite{perali2004bcs,perali2004quantitative,perali2012anomalous,innocenti2010resonant,innocenti2010shape,innocenti2013isotope,bianconi2013shape,bianconi2015lifshitz,bianconi2015superconductivity,bianconi2020superconductivity,valletta1997electronic,valletta1997t,mazziotti2017possible}]
 or the Dirac equation [\onlinecite{mazziottimultigap}] for the particular nanoscale superlattice in the real space. This provides 
 the line-shape of the $T_c$ amplification by Fano-Feshbach resonance due
to configuration interaction between different pairing channels in the two-gap superconductivity
 in artificial superlattices of quantum wires and quantum wells which is missing in the single-gap BCS theory.  
The $T_c$ amplification as function of Lifshitz parameter shows the typical Fano line-shape resonance with a dip 
at the Fano antiresonance near $\eta$=0 and a maximum amplification of $T_c$ for $\eta$ in the range $1-1.5$.
In  superlattices of quantum wells the topological Lifshitz transition from a closed 3D Fermi surface to an open
corrugated cylindrical Fermi surface, called opening a neck, occurs at $\eta$=1 with only a singular critical point in the k-space,
which is not enough to produce a peak in the density of states.
Recently it has been reported that  it is possible to get a further amplification of the critical temperature 
$T_c$ for superconductivity in a two-band two-gap superconductor  by including spin orbit interaction [\onlinecite{mazziottimultigap}].

In this scenario the quantum calculation of the energy dependent pair-transfer exchange  term requires
the solution of the Dirac equation for the superlattice. The RSOC was proposed to be controlled 
by the internal electric field $E$ at the interface between nanometer thick metallic layers intercalated 
by insulating layers forming the heterostructure at atomic limit.
In  the presence of RSOC  the $T_c$ shows a maximum at the topological Lifshitz transition from a $torus$ to a 
corrugated $cylinder$ associated with an unconventional higher order van Hove singularity (VHS) associated with a circular line of critical points 
and a sharp peak in the DOS increasing with  RSOC strength[\onlinecite{mazziottimultigap}].  

The maximum critical temperature in the curves $T_c$ versus $\eta$ occurs at $\eta$ near 1 and electron-phonon
coupling ($g=0.4$) in the second subband and weak coupling ($g$=0.1) in the first subband  [\onlinecite{mazziottimultigap}]. 
However the role of the electron-phonon coupling in the small appearing Fermi surface in the Fano-Feshbach 
resonance remains elusive. In fact  large variations of the electron-phonon interactions with the chemical potential,
due to the emergence of a Kohn anomaly at particular spots in the  k-space,  was experimentally determined in the archetypal case 
of two-gap layered alloy $Mg_{1-x}Al_{x}B_2$ by Simonelli et al. [\onlinecite{di2002amplification,simonelli2009isotope}].
While it is known that the critical temperature as a function of doping and  electron-phonon interaction
 in several families of high-temperature superconductors  shows a 3D phase diagram 
[\onlinecite{agrestini2001high,bianconi2001quantum,agrestini2003strain,he2018rapid}], 
the superconducting $dome$ remains elusive.

In this work we fill this gap  by first-principle calculations of the resonant superconducting domes for the two-gap resonant superconductivity  in the artificial superlattice introduced above.  We include   the energy dependent intra-band electron-phonon coupling for electrons in the upper subband 
forming the appearing Fermi surface. We report the superconducting dome in a 3D color-plot for $T_c$ as a function  of  two  parameters, i.e.   (i) the Lifshitz parameter measuring the position of the chemical potential relative to the Lifshitz transition  
and (ii) the strength of the electron-phonon interaction in the upper subband in the absence and in the presence of increasing 
RSOC. 
As we shell see the top of the superconducting dome is given by the Fano-Feshbach resonance driven by quantum configuration interaction between open and closed pairing channels, i.e., between a first gap in the BCS regime, resonating with a second gap in the BCS-BEC crossover regime. We show that the maximum $T_c$ increases with increasing RSOC interaction and we propose the possible pathway in the ($\eta$) and $g$ plain due joint variation of doping and strain inspired by experimental data on Al doped $MgB_2$

\section{Results and discussion}
\subsection{Fano-Feshbach resonance}

\begin{figure}
	\centering
	\includegraphics[scale=0.7]{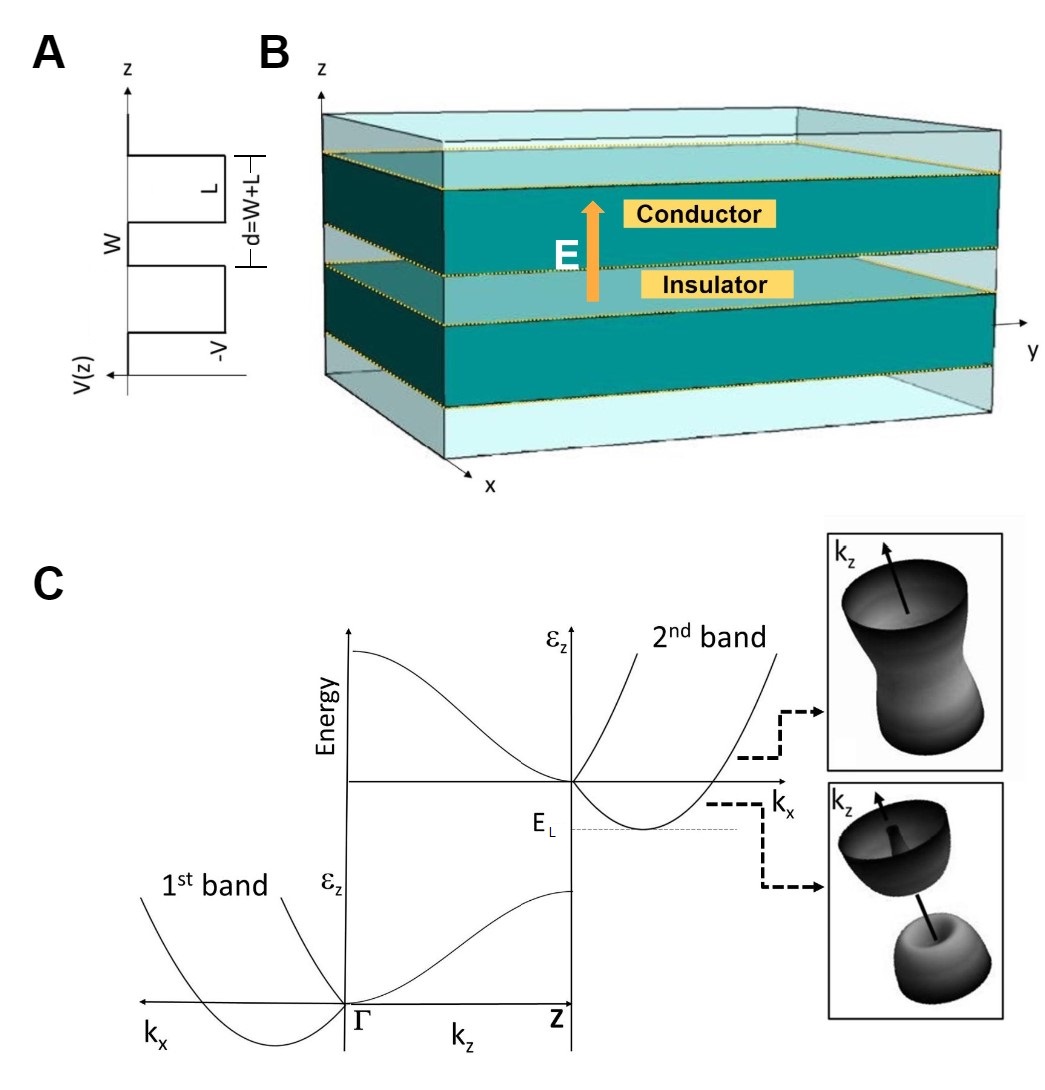} 
		\caption{Pictorial view of the 3 nm superlattice of quantum wells studied in this work. The panel A shows the periodic potential 
	along the z-direction where the electrons are confined in the metallic layers of thickness $L=2.3\ nm$ 
	(dark green layers in panel B) separated by the potential barrier of width $W=0.7\ nm$  made of insulating 
	layers (light green in panel B). In this nanoscale superlattice an internal electric-field, \textbf{E}, is present which makes possible a Rashba coupling for the electron gas in the thin metallic layers.
    Panel C:  the dispersion of the first two superlattice subbands are shown along $k_x$ and $k_z$ in the presence of the RSOC. The RSOC shifts the bottom of the second subband at $E_L$ by an amount $E_0$  defined in the main text. In the panels on the right we have plotted the Fermi surfaces for the negative helicity subband above and below the anomalous van Hove singularity due to the spin-orbit coupling. The intensity of the gap, $\Delta_2$, in the second subband increases going from dark to light colour.}

\label{fig:structure}
\end{figure}

 The proposed heterostructure has  periodicity $d=30$ \AA and is made of  alternating superconducting layers of thickness $ L = 23$ \AA   and insulating  layers of thickness $ W = 7$ \AA. The insulating layers are realized   with a potential barrier $V=0.5\ eV$ as shown in Fig. {\ref{fig:structure}}. 
This is a selected a prototype of an artificial superlattice of nanoscale quantum wells made of a conventional superconductor intercalated by insulating layers where the tunneling of the electrons through the barrier is of the order of the energy cut-off of the pairing interaction. The particular case discussed in this work provides a typical example which grabs the essential physics for practical realizations of  artificial nanoscale heterostructures made of quantum superlattices with period in the range of between 1.5 and 4 nm.

Whereas along the stacking direction z the electrons experience a periodic potential, they are free in the xy plane. As a result, the non-interacting single-particle  wave functions are given by
\begin{equation}
\psi_{n\mathbf{k}\lambda}\left(\mathbf{r}\right)=\varphi_{nk_{z}}\left(z\right)\frac{e^{i\mathbf{k}_{\parallel}\cdot\mathrm{r}_{\parallel}}}{\sqrt{\mathcal{A}}}\boldsymbol{\chi}_{\lambda}\left(\theta_{{\bf k}_{\parallel}} \right),\label{eq:wavefunction}
\end{equation}
where the wave vector components $\mathbf{k=}\left(k_{x},k_{y},k_{z}\right)\equiv\left(\mathbf{k}_{\parallel},k_{z}\right)$
label plane waves in the xy plane of area $\mathcal{A}$ and the Bloch
functions $\varphi_{nk_{z}}\left(z\right)$ along the z axis, $n$
being a subband index and $\lambda=\pm 1$  the helicity index. In the above $\theta_{{\bf k}_{\parallel}}$ is defined as the angle between the $\mathbf{k}_{\parallel}$ wavevector and the $k_x$ axis. The functions $\varphi_{nk_{z}}\left(z\right)$
and the corresponding eigenvalues are obtained by imposing the continuity of the wave function and its first derivative at the discontinuity points of the potential. Finally,  ${\boldsymbol \chi}_{\lambda}(\theta_{{\bf k}_{\parallel}})$ are the spinors which are the eigenstates of the Rashba spin-orbit coupling. In the following the attractive interaction will be projected in the Rashba eigenstates as done in Ref.[\onlinecite{gor2001superconducting}]. Then, in this $3$ $nm$ superlattice  the quantum size effects split the electronic spectrum in  $n$-subbands. 
In the following we focus in two-gap superconductivity where the chemical potential is tuned by strain or charge density 
near the band edge of the second, $n=2$, subband
[\onlinecite{valletta1997electronic,bianconi1998superconductivity,bianconi2005feshbach,innocenti2010resonant,innocenti2010shape,shanenko2012atypical,bianconi2014shape,jarlborg2016breakdown,mazziotti2017possible,cariglia2016shape,innocenti2010shape,innocenti2010resonant,guidini2014band,guidini2016bcs,perali1996gap,perali1997isotope,perali2004bcs,perali2004quantitative,perali2012anomalous,doria2016multigap,salasnich2019screening,shanenko2006size}] 
including the Rashba spin orbit  coupling [\onlinecite{gor2001superconducting,rashba1960properties,cappelluti2007topological,caprara2012intrinsic,caprara2014inhomogeneous,brosco2017anisotropy,mazziotti2018majorana,mazziottimultigap}].

In agreement with Ref.[\onlinecite{mazziottimultigap}] the Rashba coupling constant  is described by the following equation:
\begin{equation}
\alpha=2\frac{\hbar^2}{2m} \frac{2\pi}{d} \alpha_{SO},
\end{equation} 
where $\alpha_{SO}$ is a dimensionless parameter that describes the strength of the Rashba coupling in units of the 
modulation parameter of the superlattice $d$ and in the numerical evaluation we will use units such that the effective mass $m$ will be fixed to one.
Then, it is possible to change the relative intensity of the Rashba coupling 
by changing the periodicity of the superlattice.

The spin-orbit coupling induces a Lifshitz
transition [\onlinecite{mazziottimultigap}] associated with a particular unconventional van Hove 
singular point [\onlinecite{volovik2017lifshitz,volovik2017topological,volovik2018exotic}] 
in the negative helicity states (we recall that the RSOC breaks the spin degeneracy yielding two helicity states at fixed wavevector $\mathbf{k}$), forming a circle whose radius increases with the RSOC  strength. 
Notice that the RSOC shifts the bottom edge $E_L$ by an amount $E_0=-(m\alpha^2)/(2\hbar^2)$.
When the Lifshitz parameter, defined in Eq.(\ref{eq:7.1}),  will be used in the presence of RSOC, it is understood that the bottom edge $E_L$ is shifted by the above amount $E_0$.
The  amplification of the gaps and  $T_c$ increases when  the chemical potential 
is tuned around an unconventional Lifshitz transition. Therefore in addition to the Cooper coupling, 
we take into account the key role of the Fano-Feshbach due to configuration interaction 
 in the equation of the gap by  evaluating  the exchange interaction
between pairs of singlets in subbands with different quantum numbers and different helicity. 

 The two-gap equations for the superlattice need to include the contact exchange interaction connecting pairs in different bands with different helicity of the pairing. The pair $\{(n, {\bf k}, \lambda),\ (n,- {\bf k}, \lambda)\}$ can be scattered into the pair  $\{(n', {\bf k}', \nu),\ (n',- {\bf k}', \nu)\}$ where $n$ and $n'$ are the band indices, $\textbf{k}$ and $\textbf{k}'$ are the wave-vectors and $\lambda$ and $\nu$ are the helicity indices. By recalling  the results of previous work (see Ref. [\onlinecite{mazziottimultigap}]), the zero-temperature two-gap equations read:
\begin{eqnarray}
\Delta_{\lambda,n} ({\mathbf k})&=&\lambda e^{i\theta_{{\mathbf k}_\parallel}} \Delta_n (k_z),\label{eq:11}\\ 
\Delta_n (k_z)&=&-\frac{U_0}{2}\sum'_{n',k'_z} I_{nk_z,n'k'_z} \Delta_{n'}(k'_z)  \sum'_{\nu,{\mathbf k'}_{\parallel}} \frac{1}{2E_{n', \nu, {\mathbf k'}}},
\nonumber
\end{eqnarray}
where the quasiparticle energy, in the superconducting state, is
\begin{equation}
E_{n',\nu,{\mathbf k'}} =\sqrt{(\varepsilon_{\nu {\bf k'}_{\parallel}}+\varepsilon_{n'k'_z}-E_F)^2+|\Delta_{n'}(k'_z)|^2},
\label{eq:12}
\end{equation}
being understood that $\mathbf{k'}=\left(k'_{x},k'_{y},k_{z}\right)\equiv\left(\mathbf{k'}_{\parallel},k'_{z}\right)$.
The presence of RSOC affects only the in-plane part of the wave functions: hence, it is convenient to indicate the eigenvalues of the single-particle non-interacting Hamiltonian with the sum $\varepsilon_{\nu {\bf k'}_{\parallel}}+\varepsilon_{n'k'_z}$, where $\varepsilon_{\nu {\bf k'}_{\parallel}}={\hbar\bf k'}^2_{\parallel}/(2m)+\lambda \alpha { k'}_{\parallel}$ and $\varepsilon_{n'k'_z}$   is given by the numerical solution of the Kroenig-Penney like problem in the z direction.
The primed sums in Eq.(\ref{eq:12}) indicate that we consider only   the pairs whose energies differ from the Fermi energy less than the renormalized cut-off Debye energy, i.e.  $|\varepsilon_{\nu {\bf k'}_{\parallel}}+\varepsilon_{n'k'_z}-E_F|<\omega_0$.
Finally, in Eq.(\ref{eq:12}) the important  exchange integral $I_{nk_z,n'k'_z}$, which carries the information of the motion along the z axis, reads
\begin{equation} 
I_{nk_z,n'k'_z}= \frac{2\pi}{L^2_\parallel}\int  |\varphi_{n,k_z}(z)|^2|\varphi_{n',k'_z}(z)|^2 dz.
\label{eq:13}
\end{equation}

One notices in  Eq.(\ref{eq:11})  that the pairing potential depends linearly on the helicity index  $\lambda$  and through a phase factor on the in-plane momentum, whereas  the quantity $\Delta_n(k_z)$ does not depend explicitly on both in-plane momentum and helicity [\onlinecite{gor2001superconducting}].
To obtain the critical temperature we need to consider the finite-temperature version of the two-gap equations of Eq.(\ref{eq:12}), which is 
\begin{equation}
\Delta_n(k_z)=-\frac{U_0}{2}\sum'_{n',k'_z} I_{nk_z,n'k'_z} \Delta_{n'}(k'_z)  \sum'_{\nu {\mathbf k'}_{\parallel}} \frac{\tanh (\frac{\beta E_{n',\nu,{\mathbf k'}}}{2})}{2E_{n',\nu,{\mathbf k'}}}.
\label{eq:83}
\end{equation}
The critical temperature is then obtained in the standard way by taking the limit $T\rightarrow T_c$, $\Delta_n(k_z)$ close to zero in the above. More precisely, thanks to the matrix structure in the subband indices of the exchange integral, we get a linear system for the unknowns $\Delta_n(k_z)$, where we confine to   $n=1$ and $n=2$. For $T=T_c$, the maximum eigenvalues of the matrix of the system is equal to 1 and this is the condition we seek when we solve for $T_c$. It is precisely this matrix structure which takes into account the interference between the electronic  wave functions in the different subbands.
The possibility of varying each matrix term allows us to study the superconducting phase when different condensates coexist
 in different coupling regimes  and when it is possible to increase the critical temperature reaching  high $T_c$ superconductivity
 in the weak coupling regime with optimal choice of the nanoscale lattice parameters of particular metallic heterostructure 
 at atomic limit [\onlinecite{mazziottimultigap},\onlinecite{mazziotti2017possible},\onlinecite{mazziotti2021resonant},\onlinecite{mazziotti2021room}].
 \begin{figure}
	\centering
	\includegraphics[scale=1.4]{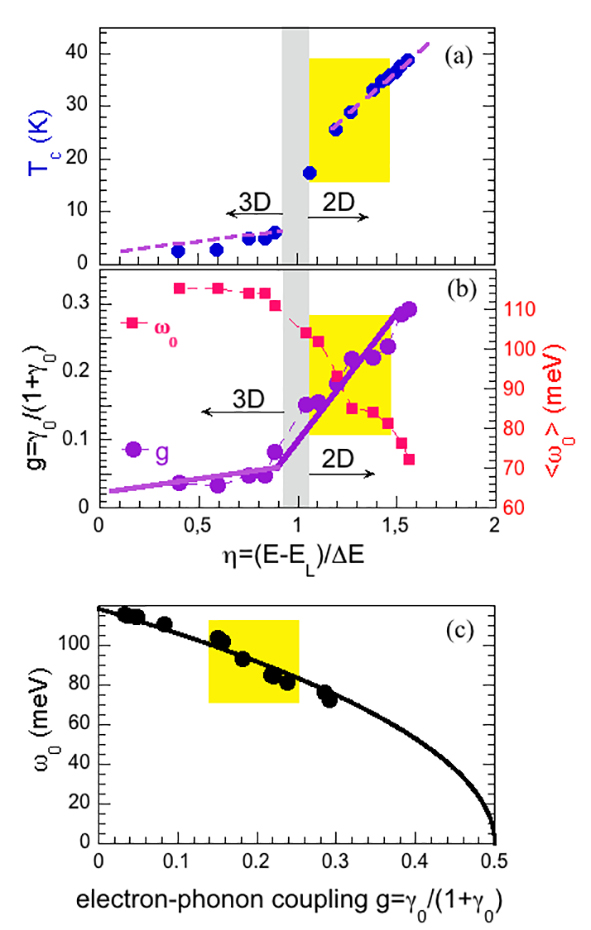}
	\caption{Panel (a) shows the superconducting critical temperature of $Mg_{1-x}Al_{x}B_2$ as a function of the Lifshitz
	 parameter  $\eta$ where $E_L$ is the energy of the chemical potential for the appearing of the $\sigma$ Fermi surface
	 and $\Delta E$ is the dispersion of the $\sigma$ band in the transversal $z$ direction [\onlinecite{simonelli2009isotope}]. 
	 Panel(b) shows the energy of the $E_{2g}$ phonon-mode, $\omega_0$, in $Mg_{1-x} Al_{x} B_2$, measured by 
	 Raman spectroscopy  [\onlinecite{simonelli2009isotope}] as a function of the Lifshitz parameter $\eta$. The yellow regions
	  indicate the phase-separation regime near the 2D-3D Lifshitz topological transition observed
	  experimentally[\onlinecite{simonelli2009isotope}, \onlinecite{agrestini2004substitution}] predicted in Ref,
	  [\onlinecite{kugel2005phase,kugel2008two,bianconi2015intrinsic,kugel2008model}]. Panel (c) shows the 
	  experimental data $\omega_0$ as a function of $g$ fitted with the Migdal relation, 
	  $\omega_0=\tilde{\omega}_0 \sqrt{1-2g}$, where $\tilde{\omega}_0=120\ meV$.}
	  \label{fig:2}
\end{figure}

To analyze the effect of increasing RSOC strength, we consider three cases 
with  $\alpha_{SO}=0.0,  \  0.4, \ 0.7$. For each of the three cases, 
we calculate $T_c$ as a function of the Lifshitz parameter $\eta$ at a fixed renormalized  electron-phonon coupling $g$.
The latter is given by  g=$\gamma_0/(1+\gamma_0)$ and is varied  in the 
range $0<g<0.5$ for $0<\gamma_0<1$, where $\gamma_0=\Gamma/\omega_0$ has
been obtained in Ref.[\onlinecite{simonelli2009isotope}] by measuring  the width of the Raman line $\Gamma$, and the phonon energy 
$\omega_0$. 

We have been inspired here by the data available for the archetypal case of natural layered two-gap superconductor, the doped 
$MgB_2$ [\onlinecite{bianconi2001superconductor}] with chemical formula $Mg_{1-x}A_{x}B_2$ with A=Al or Sc, 
where the $T_c$(x) and $T_c$($\eta$) have been measured [\onlinecite{di2002amplification,agrestini2004substitution}].
In the doped  $MgB_2$ the first Fermi surface is made of boron $\pi$ orbitals and the second appearing Fermi surface is made 
of quasi two-dimensional boron $\sigma$ orbitals. 
The critical temperature $T_c$  in $Mg_{1-x}Al_{x}B_2$ as a function of the Lifshitz parameter $\eta$ is 
plotted in panel (a) of Fig. \ref{fig:2}. In fact the Al(3+) dopants replace Mg(2+) ions in the spacer atomic layers and therefore 
tune the charge density in the boron layers via the charge transfer from the spacer layers to the superconducting boron layers
(due the different charge in the Al dopant ion and $Mg$ ion). Band structure calculations have been 
used to get the transverse hopping energy $\Delta E$ between boron $\sigma$ two-dimensional 
orbitals [\onlinecite{di2002amplification,agrestini2004substitution}] needed for the calculation of $\eta$. 
 Al doping in $Mg_{1-x} Al_xB_2$ induces the variation of the electron-phonon coupling for the carriers in the appearing $\sigma$ band 
which has been measured by Raman spectroscopy in Ref. [\onlinecite{simonelli2009isotope}].
Panel (b) of Fig. \ref{fig:2} shows the softening of the energy $\omega_0$ of the $E_{2g}$ phonon mode 
and the variation of the reduced electron-phonon coupling $g$ as a function of $\eta$ in in $Mg_{1-x} Al_xB_2$.  
In the case of $Mg_{1-x}Al_{x}B_2$ it is necessary  to take into account the presence of nanoscale phase separation
near the Lifshitz transition [\onlinecite{palmisano2008controlling}] which appears in the yellow area in Fig. \ref{fig:2}, where the data indicate 
the averaged values of the two split Raman lines. 
In fact while for a non interacting Fermi gas the topological 
Lifshitz transitions are of 2.5 order, here for interacting particles in the appearing of a new Fermi surface the Lifshitz transition becomes 
a first order transition with phase separation 
[\onlinecite{kugel2005phase,kugel2008two,bianconi2015intrinsic,kugel2008model}] and critical 
opalescence 
[\onlinecite{bianconi2001quantum,bianconi2000stripe,yamaji2006condensed,misawa2006quantum,yamaji2007quantum}].
 
The panel (c) of Fig. \ref{fig:2} shows the experimental softening of the energy $\omega_0$ of the $E_{2g}$ phonon mode in
$Mg_{1-x}Al_{x}B_2$ [\onlinecite{simonelli2009isotope}] as a function of the electron-phonon coupling $g$. 
The figure shows that the data can be fitted with the Migdal relation  [\onlinecite{migdal1958interaction}], between the renormalized 
cut-off energy, $\omega_0$, and $g$, where the bare cut-off energy is given by $\tilde{\omega}_0$:
\begin{equation}
\label{eq:7.2}
\omega_0=\tilde{\omega}_0 \sqrt{1-2g}.
\end{equation}
This relation (\ref{eq:7.2}) shows that by tuning the chemical potential near the Kohn anomaly  [\onlinecite{agrestini2004substitution}] 
 the phonon energy approaches zero and the electron-phonon coupling approaches $g=0.5$. 
The fitted Migdal relation between $\omega_0$ and $g$ is shown in panel (c) of Fig. \ref{fig:2} 
 for the material design  of nanoscale artificial superlattices of quantum wells. 
 \begin{figure}
	\centering
	\includegraphics[scale=1.0]{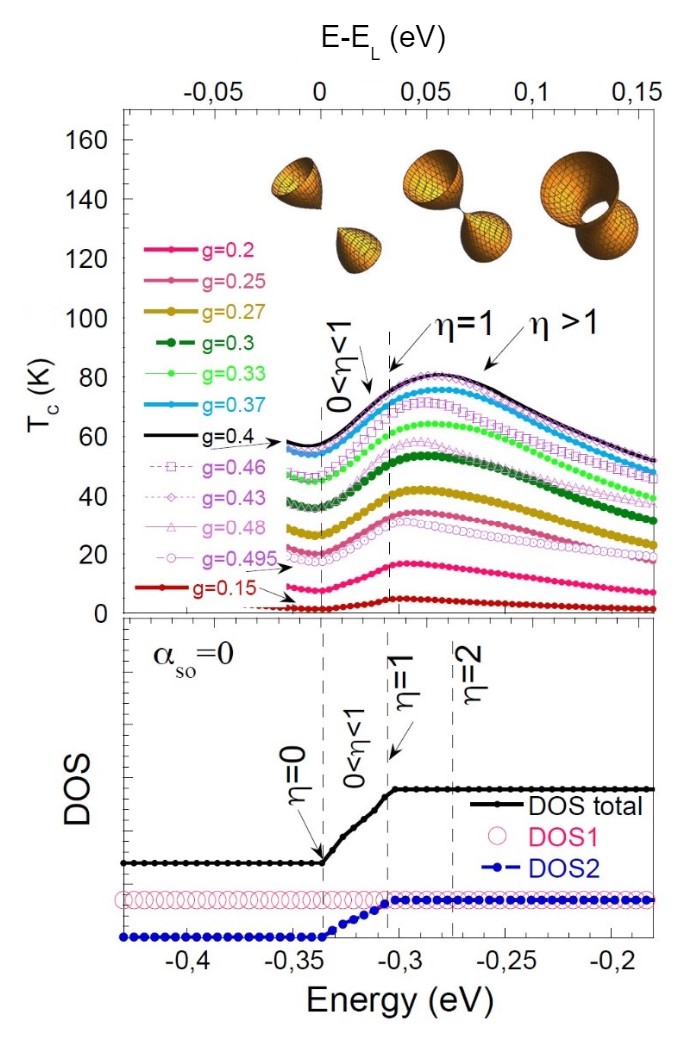} 
	\caption{ The superconducting critical temperature as function of the chemical potential measured relative to the top  of the periodic potential barrier (taken as the zero energy) and the density of states, DOS, of the normal phase in lower panel  of an artificial superlattice of quantum wells without RSOC i.e., $\alpha_{SO}=0$. On the upper part of the figure we show  the Fermi surface topology for three typical values of the Lifshitz parameter, $\eta\gtreqless1$ defined by the Eq.\ref{eq:7.1}.  $T_c$ is plotted as a function of the energy of the chemical potential  for many values of the electron-phonon coupling $g$ in the range $[0:0.5]$ where the energy cut-off is related to $g$ following the Migdal relation (Eq.\ref{eq:7.2}) as in 
 panel (c) of Fig. \ref{fig:2}. The critical temperature  reaches a maximum value equal to $80\ K$ for $g=0.4$. In the lower panel we plot the partial DOS for the first subband (pink dots) and for the second subband (blue dots)  with the total DOS (black dots) as a  function of the energy, showing the typical pattern of a superlattice of interacting layers.}
\label{fig:7.3}
\end{figure}

\begin{figure}
	\centering
	\includegraphics[scale=1.1]{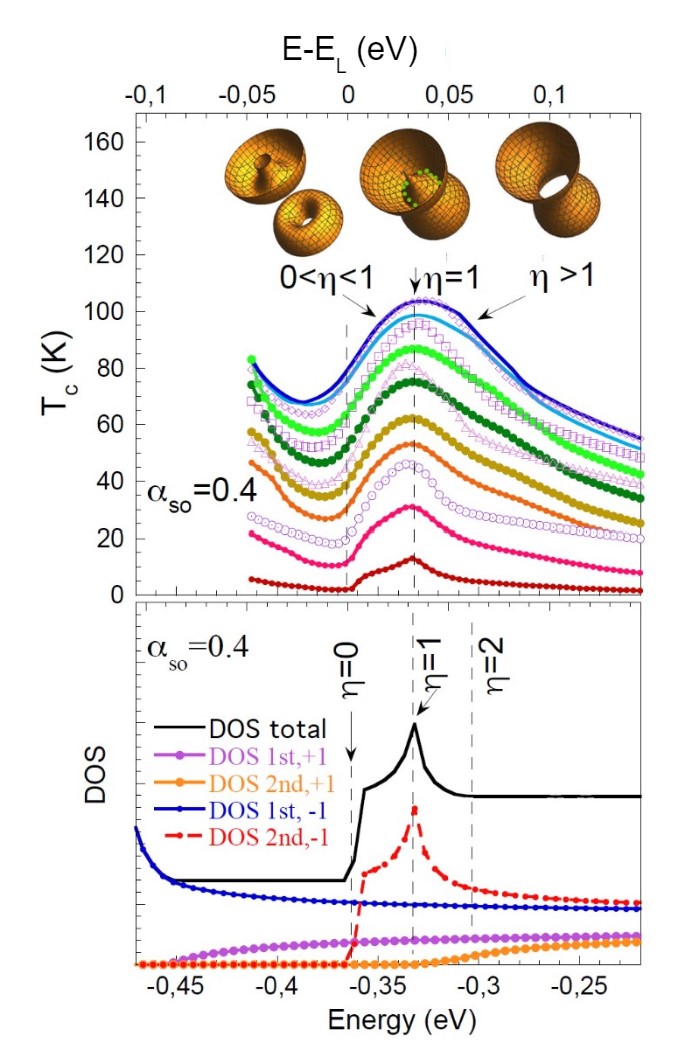} 
	\caption{  The superconducting critical temperature, panel (a), as function of the chemical potential measured relative to the top 
	of the periodic potential barrier (taken as the zero energy level) and the density of states, DOS, panel (b), of  the normal phase for 
	an artificial superlattice of quantum wells with an internal electric field, giving a spin orbit coupling, $\alpha_{SO}=0.4$. 
	The Fermi surfaces for three values of the Lifshitz parameter, $\eta\gtreqless1$  defined by the Eq.\ref{eq:7.1} are shown 
	in the upper part of the figure to show the topological Lifshitz transition of the appearing new Fermi surface from a torus to
	 a corrugated cylinder for $\eta > 1$.}	
\label{fig:7.4}
\end{figure}

Fig. {\ref{fig:7.3}} shows the calculated superconducting temperature $T_c$ as a function of the chemical potential for 
the superlattice shown in Fig. 1 and the total and partial density of states, DOS, of the normal phase for $\alpha_{SO}=0$. 
For $\alpha_{SO}=0$ the total DOS shows the step-like behavior typical of the superlattice of quantum wells at the appearing 
of the 3D Fermi surface and the flat density DOS where the Fermi surface is a corrugated cylinder above the Lifshitz 
transition of the type opening a neck.
The  panel (a) of the  Fig. {\ref{fig:7.3}} shows the critical temperature as a function of the  chemical potential for different values of the electron-phonon coupling $g$ and different phonon cut-off energy $\omega_0$, 
which are related by the phenomenological Migdal relation shown in panel (c) of Fig. \ref{fig:2}.
In the lower part of the Fig. {\ref{fig:7.3}} we report the partial DOS for the first two subbands and the total DOS of the superlattice. 
The critical temperature shows a dip when the chemical potential is tuned at the band-edge 
of the second subband ($\eta=0$), due to the negative interference effect typical of a Fano antiresonance between 
the two pairing channels.
We observe at $\eta=0$ a topological Lifshitz transition of the first type, with the appearing of a new closed 3D Fermi surface
[\onlinecite{bianconi1994fermi,bianconi2000strain,bianconi2001superconductor,bianconi2005feshbach,bianconi2015superconductivity,bianconi2020superconductivity,innocenti2013isotope,perali2012anomalous,valletta1997t}]. 
At $\eta=1$, we have a second Lifshitz transition of the second type,
called {\sl opening a neck}, where the Fermi surface changes the topology from 3D to a 2D corrugated cylinder. 
At this VHS there is only a singular point which is not enough to give a peak in the DOS, but it gives a peak
in the critical temperature. For values of $\eta > 1$ the Fermi surface becomes a 2D corrugated cylinder.
This figure shows that the maximum of the critical temperature due to the Fano-Feshbach resonance occurs 
where the Lifshitz parameter is in  the range  $\eta=1.5$, i.e. well above the VHS associated to an isolated singularity. 
The results shows that in  the absence of the RSOC it is possible to reach a maximum of the critical temperature  
of $80\ K$ with a selected electron-phonon parameter $g=0.4$, and about $40\ K$ for $g=0.3$ like in $MgB_2$.

Fig. {\ref{fig:7.4}} shows the curves $T_c$  versus the chemical potential  in the presence of RSOC $\alpha_{SO}=0.4$.
The Fermi surfaces at three characteristic values of the Lifshitz parameter, for $\eta<1$, 
 at  $\eta=1$ and for $\eta>1$ showing the topological Lifshitz transition in the upper part of the figure.
 In panel (a) of Fig. \ref{fig:7.4} we plot $T_c$ as a function of the chemical potential for different values of the superconducting electron-phonon coupling $g$ in the range $[0:0.5]$ where the energy cut-off is related to $g$ following the Migdal relation (Eq.\ref{eq:7.2}) following the phenomenological relation shown in panel (c) of Fig. \ref{fig:2}. 
 The critical temperature increases with $g$ for $g<0.4$ where it reaches the maximum value $T_c =120\ K$. 
In panel (b) we plot the partial DOS for the first and for the second subband, for two values of the helicity and the total DOS (black dots) as a function of the energy. For positive helicity the DOS shows a trend very similar to that observed in the absence of RSOC.
In the case of negative helicity we observe a sharp peak in the DOS which is linked to the unconventional Lifshitz transition of the Fermi surface  topology from a torus to a corrugated cylinder.

For negative helicity, $ \lambda = -1 $, the DOS shows a peak where the Lifshitz parameter is equal to $E_0 / \omega_0$, where
$E_0=-(m\alpha^2)/(2\hbar^2)$ is the Rashba energy shift. As it was shown in Ref.[\onlinecite{mazziottimultigap}], the peak
in the DOS is linked to an unusual variation in the topology of the Fermi surface (see the top panel in the Fig. {\ref{fig:7.4}). In
particular, when the chemical potential reaches the second Lifshitz transition, an unconventional VHS is 
observed where the singular nodal points on the Fermi surface form an entire circle (highlighted with a green dashed curve), which 
explains the appearance of a peak in the DOS. The radius of the circle  of singular points 
increases as the RSOC increases and this is associated with the shift of the DOS peak to lower energy
(see Ref.[\onlinecite{mazziottimultigap}]).  The unconventional Lifshitz transition observed in the normal phase 
generates  an amplification of the maximum critical temperature that goes from $80\ K$, in the absence of RSOC, 
up to $120 \ K$ for $\alpha_{SO}=0.4$. By further increasing  the spin-orbit coupling up to $\alpha_{SO}=0.7$, shown in Fig. \ref{fig:7.5}, 
it is possible to reach at the top of the dome a critical temperature of $160\ K$, as in cuprate perovskite superconductors. 

\begin{figure}
	\centering
	\includegraphics[scale=1.0]{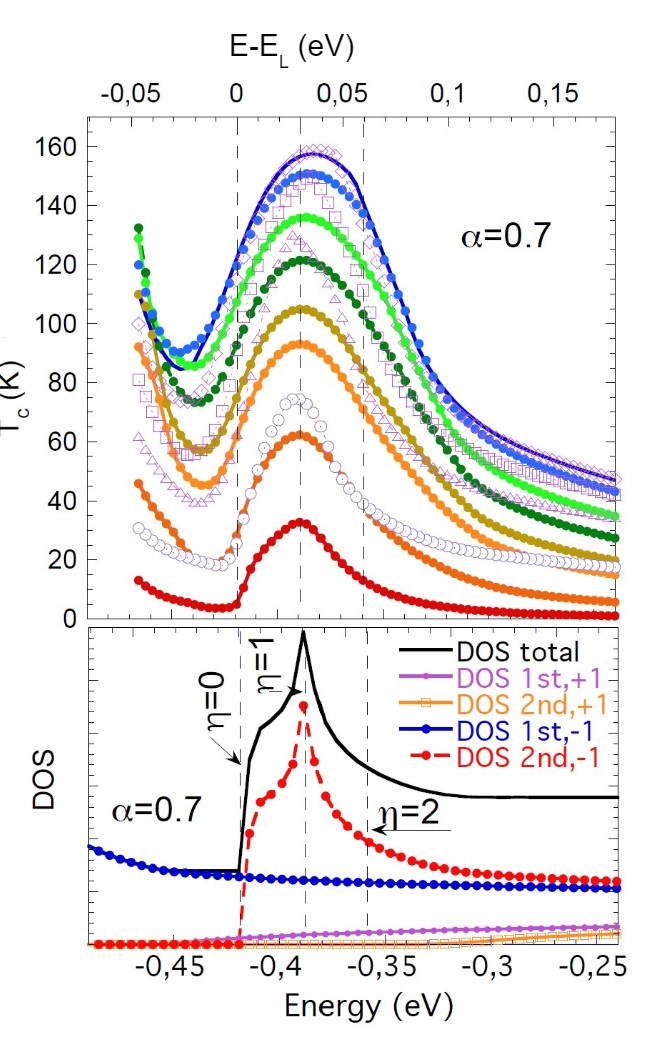} 
	\caption{The superconducting critical temperature and the density of states, DOS, of the normal phase  of an artificial 
	superlattice of nanoscale layers (heterostructure of quantum layers) with RSOC, where $\alpha_{SO}=0.7$, as a function
	 of the chemical measured relative to the top of the potential barrier  Starting from
	 the top we plot the Fermi surfaces for three values of the Lifshitz parameter, $\eta\gtreqless1$, in the central panel we 
	 plot $T_c$ as a function of the energy for different value of the superconducting coupling $g$ in the range $[0:0.5]$, 
	 according to the Migdal relation (Eq.\ref{eq:7.2}). The critical temperature increases with $g$, reaches a maximum value 
	 equal to $160\ K$ and goes toward zero at g=0.5. In the lower part of the figure we plot the partial DOS for the first 
	 and for the second subband, for two values of the helicity and the total DOS (black dots) as a function of the energy. 
	 This figure shows that by increasing the intensity of the RSOC, the peak of DOS increases and 
	 with it also the maximum critical temperature.	}
\label{fig:7.5}
\end{figure}

\subsection{The superconducting dome}

It is known that in the BCS  theory for a single-gap superconductor the critical temperature depends
on the renormalized electron-phonon coupling strength $g$, with the phonon energy, 
$\omega_0$, playing the role of a pre-factor and of the energy cut-off of pairing interaction. Therefore in the BCS theories $T_c$ it is 
not dependent on the variations of the chemical potential, since the Fermi level $E_F$ is assumed to be far from band
 edges and much larger than the energy cut-off, $E_{F}$/$\omega_0\gg1$.
 
In two-gap superconductivity in artificial nanoscale superlattices near a topological Lifshitz transition the Fermi level in the 
new appearing second Fermi surface $E_{F2}$ is near the band edge $E_L$ (see  the Eq.{\ref{eq:7.1}} ) which is shifted  in the presence of RSOC.
Moreover the Fano resonance occurs  where the Lifshitz parameter is tuned in the range 
-1<$\eta$<3 where 0<$(E_{F2}-E_L)$/$\omega_0$<2.
Therefore in the two-gap superconductivity studied here the  BPV theory allows to calculate the \textit{superconducting dome} where the critical temperature depends 
on electron-phonon coupling $g$ and the Lifshitz parameter $\eta$.
Here we report  the \textit{superconducting dome} where $T_c$ is a function of either the Lifshitz parameter $\eta$ 
or the electron-phonon coupling $g$. The \textit{superconducting dome} characterizes the superconducting properties of the artificial nanoscale superlattice with a given architecture needed for material design of novel artificial high-$T_c$ superconductors.

\begin{figure}
\centering
\includegraphics[scale=1.4]{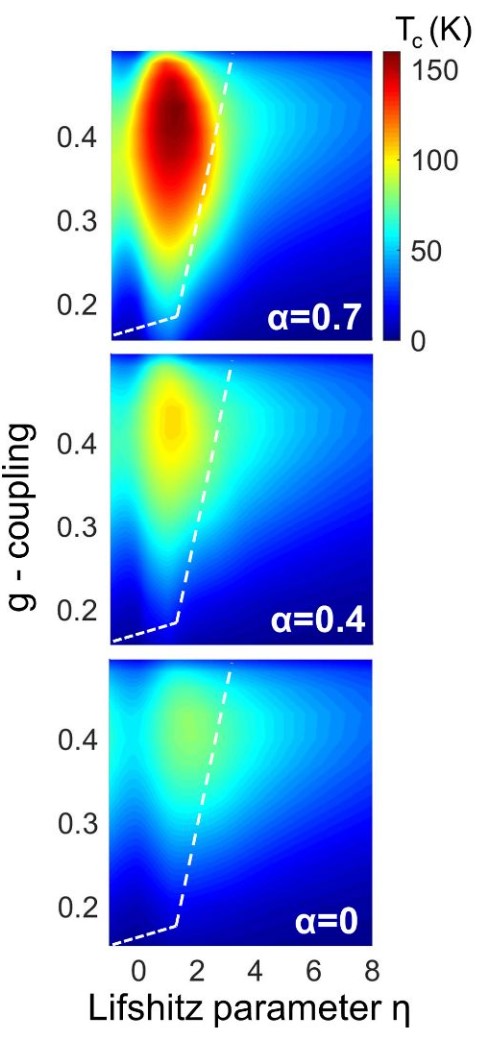}
\caption{ 
The three \textit{superconducting domes} for artificial superlattices of quantum wells for three different values of the Rashba coupling constant $\alpha_{SO} = 0, 0.4, 0.7$. The colour plots show the superconducting critical temperature as a function of the Lifshitz parameter $\eta$ measuring the position of the Fermi level $E_{F2}$ and of the electron-phonon coupling $g$ in the second subband.
The critical temperature increases with the colour going  from blue to red. 
The top of \textit{superconducting dome} occurs near $g = 0.4$ and $\eta$ about $1$. We observe a shift of the top of the \textit{superconducting dome} to lower $\eta$ values from  $\eta$=1.5 to $\eta$=1 increasing the spin-orbit coupling.
The observed marked anisotropy of the superconducting dome is due to anisotropic line-shape of the Fano resonance. 
The white dashed line indicates the pathway in the ($g$,$\eta$) plane extracted from the experimental data for  $Mg_{1-x}Al_{x}B_2$  shown in 
 panel (b) of Fig. \ref{fig:2}. In this pathway for $\eta < 1$ the electron-phonon coupling g is nearly constant around 0.1 and 0.15 while it shows a sharp change for 1<$\eta$<2.5 with a rapid increase due to a Kohnanomaly. 
}\label{fig:7.6}
\end{figure}

Using the numerical results presented in the previous subsection we show in  Fig. \ref{fig:7.6} the \textit{superconducting domes}
 as colour plots of the critical temperature $T_c$ as a function of the electron-phonon coupling $g$ and the Lifshitz
 parameter $\eta$ for three different values of spin orbit interaction RSOC $\alpha = 0, 0.4, 0.7$.
 The \textit{superconducting domes} in  Fig. \ref{fig:7.6} clearly shows that the maximum of the critical temperature is reached when the 
chemical potential is tuned at the unconventional topological Lifshitz transition of the type  \textit{opening a neck} near $\eta$=1, 
where the Fermi surface topology of the second subband  changes from a closed 3D surface to 
a corrugated cylinder for $\alpha = 0$ and from a torus to a corrugated cylinder for $\alpha = 0.4, 0.7$ with a circle of singularity 
points giving a peak in the DOS. The top of the  \textit{superconducting dome} of the critical temperature increases
 with RSOC going to $90\ K$ and $170\ K$ as the spin-orbit coupling increases from zero to 0.4 and 0.7 respectively.

Empirical 3D phase diagrams, showing \textit{superconducting domes} have been measured for high $T_c$ superconductors, beyond the standard 2D phase diagram ($T_c$ versus $doping$) introducing a second material dependent physical variable to explain experimental data of high $T_c$ superconductivity. The third axis has been proposed to be the lattice strain [\onlinecite{agrestini2001high,agrestini2003strain}], or electron-phonon  interaction [\onlinecite{bianconi2001quantum,he2018rapid}]. 
The difficulty in the interpretation of empirical 3D phase diagrams of high temperature superconductors from experimental data is due to the fact that in the experiments the variation of $T_c$  is measured for changes of chemical doping, or external pressure or  strain (micro-strain, misfit-strain) or gate-voltage  which induce a joint variation of the chemical potential and electron-phonon coupling. Therefore the interpretation of the experimental curves of  $T_c$ versus $\eta$ requires the knowledge of the response of  $T_c$ along a particular pathway in the ($\eta$,g) plane.

The \textit{superconducting dome} for artificial superlattice of quantum layers  shown in Fig. \ref{fig:7.6} has been
obtained using the phenomenological Migdal relation shown in panel (c) of Fig. \ref{fig:2} by fitting experimental data of doped $MgB_2$. Therefore we can use the phenomenological pathway in the plane ($\eta$,g) shown in panel (b) of Fig. \ref{fig:2},  which is indicated by the white dashed line in Fig. \ref{fig:7.6} to establish the theoretical prediction of the curve $T_c (\eta)$ for Al doping in $Mg_{1-x} Al_xB_2$. 
The pathway of $g$ versus $\eta$ in Fig. \ref{fig:7.6} shows low values of $g$ with a nearly flat slope for $0<\eta<1$, 
where the second Fermi surface has a closed 3D shape and  a very sharp change at the slope at $\eta=1$, with $g$ rapidly increasing in the range $1<\eta<3$,, where the second Fermi surface is a corrugated cylinder. 

In Fig. \ref{fig:7.7} we report the variation of the critical temperature $T_c$ versus $\eta$ where the superlattice of quantum wells is manipulated by doping, strain, pressure, or gate-voltage following the particular pathways in the ($\eta$,g) plane shown by the dashed line in Fig. \ref{fig:7.6}.
The  predicted $T_c$($\eta$) curve for this particular pathway shown of Fig. \ref{fig:7.7} for $\alpha = 0$ is 
compared with the experimental curve of $T_c$ versus $\eta$) extracted from data on $Mg_{1-x}Al_{x}B_2$ in Ref.[\onlinecite{simonelli2009isotope}] shown in panel (a) of Fig. \ref{fig:2}). The theory is able to reproduce the
experimental data up to $ T_ {c,MAX} = 40 \ K$ where $\eta=1.5$ and $g=0.3$


\begin{figure}
\centering
\includegraphics[scale=1.4]{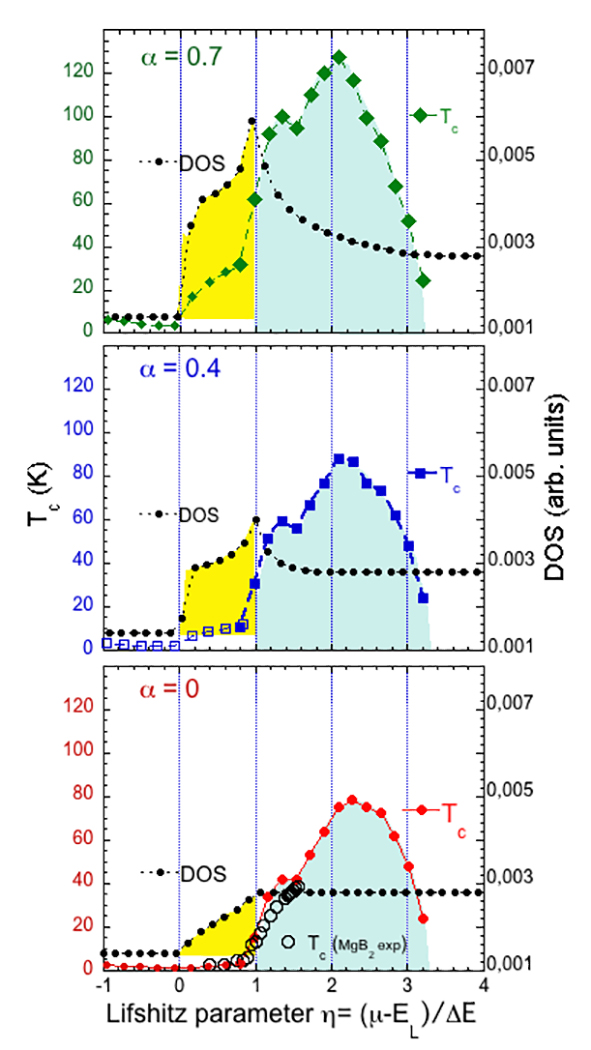}
\caption{{ 
The critical temperature $T_c$ versus $\eta$ for the 3 nm artificial superlattice of quantum wells as a function of the particular pathway
 in the ($\eta$,g) plane indicated by the dashed line in Fig. \ref{fig:7.6} where the three panels show three different cases of 
 spin orbit interaction RSOC $\alpha = 0, 0.4, 0.7$. The calculation for the case  $\alpha = 0$ is compared with experimental 
 data (open circles) measured in archetypal case of two-gap superconductivity in multi-layers 
 $Mg_{1-x} Al_xB_2$ [\onlinecite{simonelli2009isotope}] tuned in the range of the Lifshitz transition
  for \textit{opening a neck}. The variation of the density of states DOS along the same pathway is plotted  
 to show that, in the scenario presented here,  the variation of the superconducting critical temperature and the variation of the DOS 
 do not show a trivial correlation.
}}\label{fig:7.7}
\end{figure}

\section{Conclusions}

Here we have reported the calculation of the 2D superconducting dome for a 3 nm superlattice of quantum wells with  Rashba 
spin orbit coupling.  The Rashba coupling induces an unconventional Lifshitz transition associated with a VHS extended on a circumference of increasing radius with the intensity RSOC in negative helicity states [\onlinecite{mazziottimultigap}]. The complexity in the properties of the normal phase is associated with  a larger amplification of the gap and $T_c$. In particular, we have seen that the amplification increases when the chemical potential is varied around an unconventional Lifshitz transition. In this situation, in addition to the intra-band Cooper pairing, it is necessary to include the key role played by the quantum configuration interaction that appears in the mean-field equation for the self-consistency of the gap and which has to be calculated considering the exchange interactions between pairs of singlets in subbands with different quantum numbers and different helicities.
The results show that $T_c$ at the top of the \textit{superconducting dome} increases up to $170 K$ by increasing spin orbit coupling.
The Rashba spin orbit coupling can be tuned by manipulation of ferroelectricity in the artificial superlattices. In fact, 
while it is known that 3D homogeneous metals cannot exhibit ferroelectricity because static internal electric fields are screened
by conduction electrons, the coexistence of ferroelectricity and superconductivity is possible 
in inhomogeneous phases like our artificial superlattices of quantum wells if the insulating intercalated layers show ferroelectricity.
In fact the coexistence of ferroelectricity and the metallic phase has been observed in
proximity to (i) phase separation at structural phase transitions [\onlinecite{testardi1975structural}],
(ii) in metallic $LiOSO_3$ that is structurally equivalent to the
ferroelectric transition of $LiNbO_3$ [\onlinecite{boysen1994neutron}] and the coexistence is favored by electronic correlation  and phase separation near metal-insulator transitions [\onlinecite{egami1993lattice,bianconi1982multiplet,bianconi2012superconductor}]. 
In the last ten years ferroelectricity and superconductivity have been observed 
in several metallic perovskites [\onlinecite{boysen1994neutron,kolodiazhnyi2010persistence,shi2013ferroelectric,russell2019ferroelectric,jena2019new,rischau2017ferroelectric,hameed2022enhanced}] 
and a tunable spin-orbit coupling has been observed in nanoscale heterogeneous phases of matter [\onlinecite{takahashi2006local,sun2004coordination,crassous2013bifeo3,ahadi2019enhancing,gabay2017ferroelectricity,rischau2017ferroelectric}]. 
We have identified the particular pathway in the ($\eta$,g) plane of the superconducting dome  $T_c$($\eta, g$)
from experimental Raman data giving $g(\eta)$ for the natural superlattice of superconducting boron atomic layers in
$Mg_{1-x}Al_{x}B_2$ intercalated by normal Al-Mg layers. In this experimental pathway the intra-band electron-phonon coupling is weak 
near the band edge in the appearing small 3D Fermi surface and it shows a rapid increase
in the corrugated  2D cylindrical Fermi surface where it crosses a Kohn anomaly approaching the maximum $T_c$.
Moreover we have added the spin degree of freedom to be able to take into account the RSOC coupling,
hence we have solved  the Dirac equation in non-relativistic limit to determine 
the correct shape of the wave-functions needed for the calculation of energy dependent pair transfer exchange interaction. 
Therefore this work provides the quantum scenario of two-gap superconductivity near an unconventional Lifshitz
transition associated with a VHS extended on a circumference in the negative helicity states, 
whose radius increases as the RSOC increases.
Moreover here we overcome the previous theoretical approaches where the adiabatic limit was considered 
i.e., where the Fermi energy is much larger than the Rashba shift and the gap energy, 
while the interesting physics occur in a regime where the Fermi energy in the second subband
is of the same order of magnitude as the  pairing energy and the spin-orbit splitting induced by the RSOC.
Finally we have shown that by design of the heterostructure nanoscale geometry, and the choice 
of different materials for metallic layers and spacer layers, it is possible to amplify the critical temperature
by using the joint effect of the Fano resonance between the superconducting gaps,
the electron-phonon interaction near a Kohn anomaly in a corrugated cylinder Fermi surface and the RSOC
in nanoscale superlattices of quantum wells.

\section{Dedication}
This paper is dedicated to Yvette Cauchois (1908-1999) [\onlinecite{wuilleumier2003yvette}], Fig. \ref{fig:Dedication}, a pioneer of X-ray spectroscopy [\onlinecite{cauchois1964spectroscopie}] and its applications to identify localised versus delocalised electronic states in condensed matter giving white lines or absorption edges respectively [\onlinecite{cauchois1949cxvi}] which later have allowed the experimental identification of two-band, two-gap superconductors. She collaborated with Ugo Fano [\onlinecite{bianconi2003ugo}] and she made the first experiment on applied physics using synchrotron radiation in Europe [\onlinecite{1963rayonnement}].

\begin{figure}
	\centering
	\includegraphics[scale=0.22]{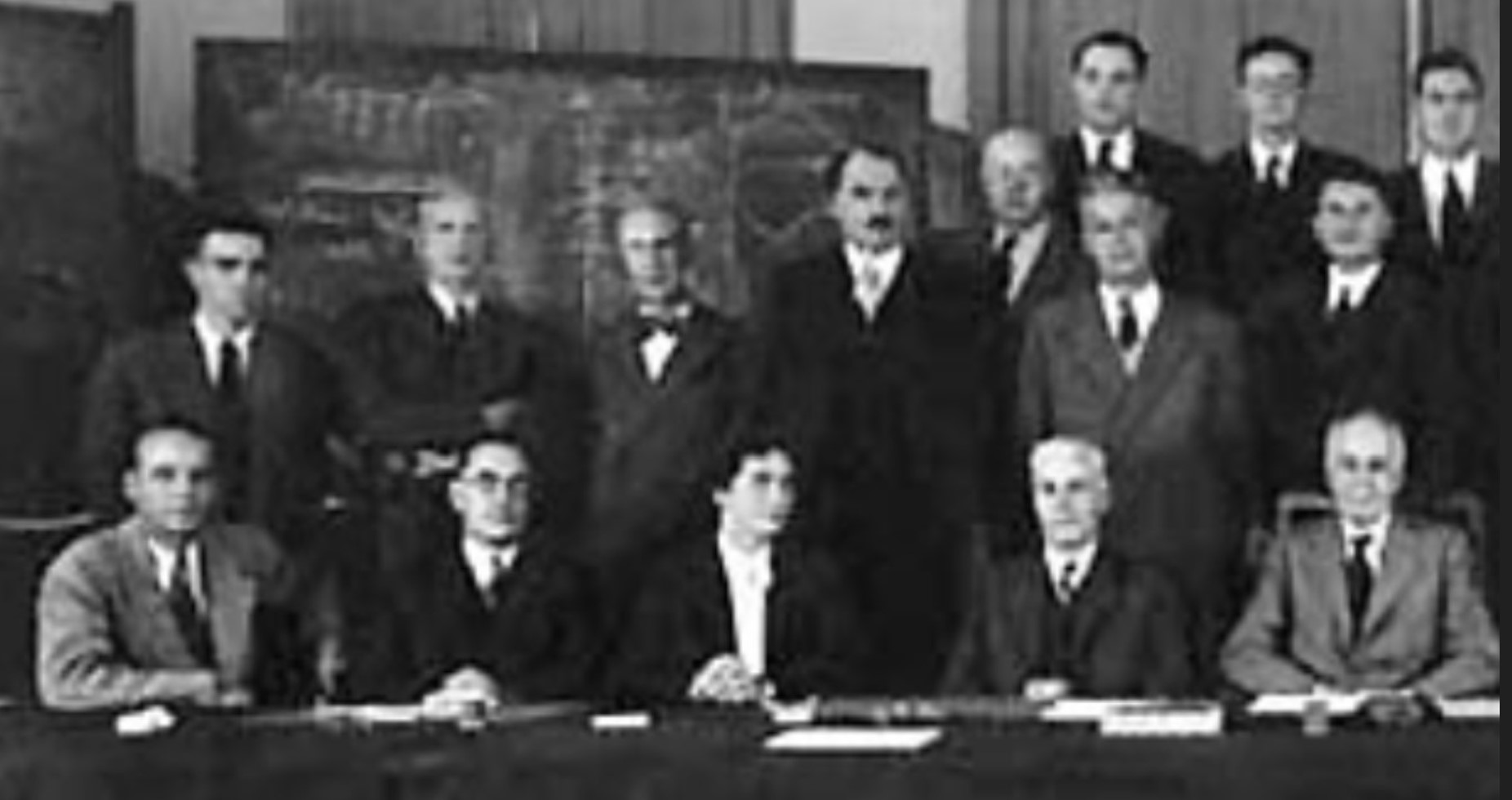}
	\caption{ Prof. Yvette Cauchois at Solvay Conference 1951, she is the third in the first row. The Picture is by G. Coopmans http://www.hilliontchernobyl.com/solvay1951.htm.	
	}\label{fig:Dedication}
\end{figure}

\begin{acknowledgments}
We acknowledge financial support of Superstripes onlus.
\end{acknowledgments}	

\input{Mazziotti2022Spin.bbl}

\end{document}

%% file: Mazziotti2022Spin.bbl
%